\documentclass[aps,12pts,amsmath,amssymb,preprint,showkeys,showpacs]{revtex4}
\usepackage{graphicx}
\usepackage{bm}
\usepackage{float}
\usepackage{mathptmx}
\usepackage{epstopdf}
\usepackage{booktabs}
\usepackage{multirow}
\usepackage{hhline}
\usepackage{verbatim}
\usepackage{subfigure}
\usepackage{soul}
\usepackage[usenames,dvipsnames]{color}

\begin{document}
\title{\textit{Ab-initio} semi-classical electronic transport in ZnSe: The role of inelastic scattering mechanisms}
\author{Anup Kumar Mandia$^{1}$, Renuka Patnaik$^{2}$, Bhaskaran Muralidharan$^{1}$},
\author{Seung Cheol Lee$^2$}
\email{seungcheol.lee@ikst.res.in}
\author{Satadeep Bhattacharjee$^{2}$}
\email{satadeep.bhattacharjee@ikst.res.in}
\affiliation{$^{1}$Indian Institute of Technology, Mumbai-400076,  India \\
$^{2}$Indo Korea Science and Technology Center, Bangalore-560065, India}

\begin{abstract} 
We present a detailed \textit{ab-initio} study of semi-classical transport in n-ZnSe using Rode's iterative method. Inclusion of  ionized impurity, piezoelectric, acoustic deformation and polar optical phonon scattering and their relative importance at low and room temperature for various n-ZnSe samples are discussed in depth. We have clearly noted that inelastic polar optical phonon scattering is the most dominant scattering mechanism over most of the temperature region. Our results are in good agreement with the experimental data for the mobility and conductivity obtained at different doping concentrations over a wider range of temperatures. Also we compare these results with the ones obtained with relaxation time approximation (RTA) which clearly demonstrate the superiority of the iterative method over RTA.
\end{abstract}
\maketitle
\section{Introduction}
Semi-classical transport models \cite{ref5,ref6,ref7,refa,refb,refc,refd,refe,reff,refg,refh} are still actively employed for transport calculations across various materials in the bulk form. Typically, they are employed to calculate the mobility, conductivity and eventually also the thermoelectric coefficients  \cite{ref5,ref6,ref7,refa,refb,refc,refd,refe,reff,refg,refh}. Most of the prevalent semi-classical simulation platforms employ the Boltzmann transport equation (BTE) within the relaxation time approximation (RTA) and well known packages have been employed. The RTA treats the relaxation time as constant or as a power law in energy and is applicable only if the relaxation time is independent of the distribution function, which is satisfied when a scattering mechanism is either elastic or isotropic \cite{ref4}. For instance, polar optical phonon (POP) scattering is neither elastic or isotropic and hence the RTA is inappropriate when one deals with POP scattering. So one has to go beyond the RTA to capture POP scattering mechanism properly. This brings a serious issue when it comes to materials which possess strong optical phonon scattering mechanisms. Methods such as the Rode iterative method \cite{ref5,ref6,ref7} or variational \cite{refc,refd} techniques are often employed to properly account for POP scattering. In this paper, we use the Rode iterative method for mobility and conductivity calculation. The objective of this work is to use ZnSe as an example to elucidate the limitations of the RTA while presenting a calculation that is based on the Rode iterative scheme solution of the BTE. \\
\indent ZnSe is a wide band gap semiconductor, which is employed for making blue light emitting diode and lasers. The electron mobility in n-ZnSe crystals was first analyzed in Ref. \cite{refi}, where the electron mobility was measured at several temperatures and it was shown that POP scattering is the most dominant scattering mechanism near the room temperature region. There have been a few more works \cite{ref6,ref7,reff,refj,refk,refl,refm} which have carried out theoretical investigations on n-ZnSe electron mobility. However, most of these calculations were based on the RTA method. Earlier works \cite{ref6,ref7,reff} had also done theoretical investigations beyond RTA for n-ZnSe electron mobility using an iterative and variational technique respectively and have shown good agreement between theory and experiment. References \cite{ref6,ref7} had investigated n-ZnSe electron mobility various scattering mechanisms over a wide temperature range with iterative technique, while Ref. \cite{reff} had  investigated the dependence on donor concentrations with variable compensation ratios with variational technique. All of these  works used parabolic or Kane model for the mobility calculation and all the constants used were taken from the experiments.\\

\indent In the present work, the electron mobility and conductivity is investigated for n-type ZnSe at different doping concentrations and temperatures with \textit{ab-initio} inputs. For our theoretical analysis, all the parameters of band structure, density of states, wave function admixture, dielectric constants, piezoelectric constants, polar optical phonon (POP) frequencies, and acoustic deformation potentials are calculated by using the density functional theory (DFT). In ZnSe, since POP scattering is the dominant mechanism and is inelastic, it modulates the electron energy, and hence a universal time constant cannot be defined, making the RTA inappropriate. We thus use the Rode iterative method with \textit{ab-initio} inputs as previously discussed in Ref. \cite{ref1}. In the Rode method, the perturbation in the distribution is obtained at low electric field, keeping only the linear term. The perturbation in the distribution function is used to calculate the mobility. Since all the required inputs are calculated by using DFT, only the crystal structure is given as input, it does not rely on experimental data. So this is the first paper who calculated n-ZnSe electron mobility with abinito band structure and inputs by using Rode method and a good agreement is obtained for different concentrations over a wide temperature range. The same approach can be applied to other new materials for which many constants are not determined experimentally.\\ 
\indent In this paper, we first compare the experimental data and RTA calculation, and establish that the RTA results underestimate the mobility by more than a fifty percent at room temperature. In the process of presenting the results here, we also give a detailed insight about the various contributions of different scattering mechanism in ZnSe at different temperatures and doping concentrations. Since all the required inputs are calculated by using density functional theory, our code can be further extended and used for mobility calculations of new materials for which many constants are not known experimentally. \\
\indent This paper is organized as follows. In the following section, we first describe in detail the methodology by first elaborating on the Rode scheme for solving the Boltzmann transport equation. We specifically hint at how this method scores over the RTA in capturing inelastic scattering processes. We also detail the computational scheme of calculating various scattering rates starting from \textit{ab-initio} simulations. In Sec. III, we describe the results in detail, where we first establish using the calculated scattering rates that the inelastic POP scattering mechanism indeed dominates over a wide temperature range in the case of n-type ZnSe. We then describe our theoretical fits in comparison with relevant experimental works at different carrier concentrations. In Sec. IV, we conclude with an outlook from this work. 
\section{Methodology}
\subsection{The Boltzmann Transport Equation} 
The Boltzmann transport equation (BTE) describes the time-evolution of the state of the charge carriers in both real and momentum co-ordinates via a distribution function, \textit{f(\textbf{r},\textbf{k},t)}, and is given by \cite{ref2,ref3,ref4}
\begin{equation}
\frac{\partial \textit{f}}{\partial t} + \textbf{v}\cdot\nabla _rf + \frac{\textbf{F}}{\hbar} \cdot\nabla _{\bf k}f= \frac{\partial \textit{f}}{\partial t}\Bigr|_{\substack{coll}},
\label{BTEE}
\end{equation}
where \textbf{v} is the carrier velocity, \textit{f} describes the probability distribution function of carriers in real and momentum space as a function of time, \textbf{F} is the applied external force, $\frac{\partial \textit{f}}{\partial t}\Bigr|_{\substack{coll}}$ represents the change in the distribution function with time due to collisions. The first term in \eqref{BTEE} represents the rate of change of the carrier distribution $\textit{f} $ with time. The second term represents the diffusion due to a gradient in the carrier density and the third term represents the change in $\textit{f} $ due to all external forces. In the presence of an electric field \textbf{E}, the external force $\textbf{F} $ is given by $e \textbf{E}$, and the BTE then becomes
\begin{equation}
\frac{\partial f}{\partial t} +\textbf{v}\cdot\nabla _rf + \frac{e\textbf{E}}{\hbar}\cdot\nabla _{\bf k}f= \frac{\partial f}{\partial t}\Bigr|_{\substack{coll}},
\label{BTE}
\end{equation}
Under steady state and spatially homogeneous conditions, the above equation can be rewritten as:
\begin{equation}
\frac{e\textbf{E}}{\hbar}\cdot\nabla _kf = -\int [ s(k,k')f(1-f') -  s(k',k)\; f'(1-f)] dk' ,
\label{BTE1}
\end{equation}
where $e$ is the electronic charge, $s(k,k')$ represents the transition rate from a state $k$ to a state $k'$.  At low electric fields, the solution to the BTE is given by the distribution function \cite{ref5,ref6,ref7}
\begin{equation} 
f(k) = f_0[\epsilon(k)] +  g(k)cos\theta,
\label{distft}
\end{equation}
where $f_0[\epsilon(k)]$ is the equilibrium distribution function, and \(\cos\theta\) is the angle between applied electric field and \(k\). Here we have neglected the higher order terms, since we are calculating mobility under low electric field conditions. Now we have to calculate the perturbation in the distribution function \(g(k)\) for calculating the low-field transport properties. The perturbation \(g(k)\) is given by \cite{ref5,ref6,ref7}
\begin{equation}
g_{k,i+1} = \frac{S_i(g_k,i)- v(k)(\frac{\partial f}{\partial z})- \frac{eE}{\hbar}(\frac{\partial f}{\partial k}) } {S_o(k)+ \frac{1}{\tau_{el}(k)}} ,
\label{pert}
\end{equation} 
Where we have considered the electric field along the z-direction. The transition rate in Eq.\ref{BTE1} can be split into two contributions: One from from inelastic (in) and other being the elastic (el) component respectively, i.e, 
$s(k,k')=s(k,k')_{in}+s(k,k')_{el}$. The expression for $S_{i}$ and $S_{o}$ which appear in Eq.\ref{pert} are given by \cite{ref7}
\begin{equation}
\ S_o(k) = \int [s_{in}(k,k')(1 - f') + s_{in}(k',k)f']dk'
\label{outsc}
\end{equation}
\begin{equation}
\ S_i(g_k,i) = \int X g_{k',i} [s_{in}(k',k)(1-f) + s_{in}(k,k')f ]dk'
\label{insc}
\end{equation}
where X is the cosine of the angle between the initial and the final wave vectors. The elastic part of the scattering rate is given by, $\frac{1}{\tau_{el}}=\int (1-X)S_{el}(k,k')dk'$.\\
As a initial guess take $g_{k,0}=0$, it will give $S_i(g_k,0)=0$. Then $g_{k,1}$ is given by 
\begin{equation}
g_{k,1} = \frac{- v(k)(\frac{\partial f}{\partial z})- \frac{eE}{\hbar}(\frac{\partial f}{\partial k}) } {S_o(k)+ \frac{1}{\tau_{el}(k)}} ,
\label{pert1}
\end{equation} 
The value of $g_{k,1}$ is used to calculate to $S_i(g_k,1)$, then $S_i(g_k,1)$ is used to calculate $g_{k,2}$.
\begin{equation}
g_{k,2} = \frac{S_i(g_k,1)- v(k)(\frac{\partial f}{\partial z})- \frac{eE}{\hbar}(\frac{\partial f}{\partial k}) } {S_o(k)+ \frac{1}{\tau_{el}(k)}} .
\label{pert2}
\end{equation} 
These steps are repeated until $g_{k,i+1}$ converges. This convergence is exponential \cite{ref7}, so normally it requires a few iteration for convergence. Typically, five iterations are required for the perturbation to converge for polar optical phonon scattering in ZnSe. Now, the calculated perturbation is used to calculate mobility and Seebeck coefficient. The effect of inelastic POP scattering is included through the terms $S_i(g)$ and $S_o$ and the effect of elastic scattering is included through the relaxation time $\frac{1}{\tau_{el}(k)}$ term. The second term in the numerator of \eqref{pert} represents the thermal driving force and the third term in the numerator represents the electrical driving force. For mobility calculations, the thermal driving force is set to zero and only the electrical driving force is considered. $\frac{1}{\tau_{el}(k)}$ is the sum of the momentum relaxation rates of all elastic scattering process, which is given by
\begin{equation}
\ \frac{1}{\tau_{el}(k)} = \frac{1}{\tau_{ii}(k)} + \frac{1}{\tau_{pz}(k)} + \frac{1}{\tau_{ac}(k)},
\label{elst}
\end{equation}
where the subscripts \textit{el},\textit{ii} , \textit{pz} and \textit{ac} are used for elastic, ionized impurity, piezoelectric and acoustic deformation potential scattering processes respectively. The carrier mobility $\mu$  is then given by \cite{ref1,ref5,ref6,ref7}
\begin{equation}
\mu = \frac{1}{3E} \frac{\int v(\epsilon) D_s(\epsilon) g(\epsilon) d\epsilon}  
{\int D_s(\epsilon) f( \epsilon )  d\epsilon },
\label{mobility}
\end{equation}
\quad
where $D_S(\epsilon)$ represents density of states. The carrier velocity is calculated directly from the \textit{ab-initio} band structure by using
\begin{equation}
\ v(k) = \frac{1}{\hbar} \frac{\partial{\epsilon}}{\partial{k}}.
\label{velocity}
\end{equation}
From these, we can evaluate the electrical conductivity given as
\begin{equation}
\ \sigma = n e \mu_e,
\label{conductivity}
\end{equation}
where $n$ is the electron carrier concentration, and $\mu_e$ is the electron mobility.  
\subsection{Computational method}
\begin{figure}
\includegraphics[width=120mm,height=100mm]{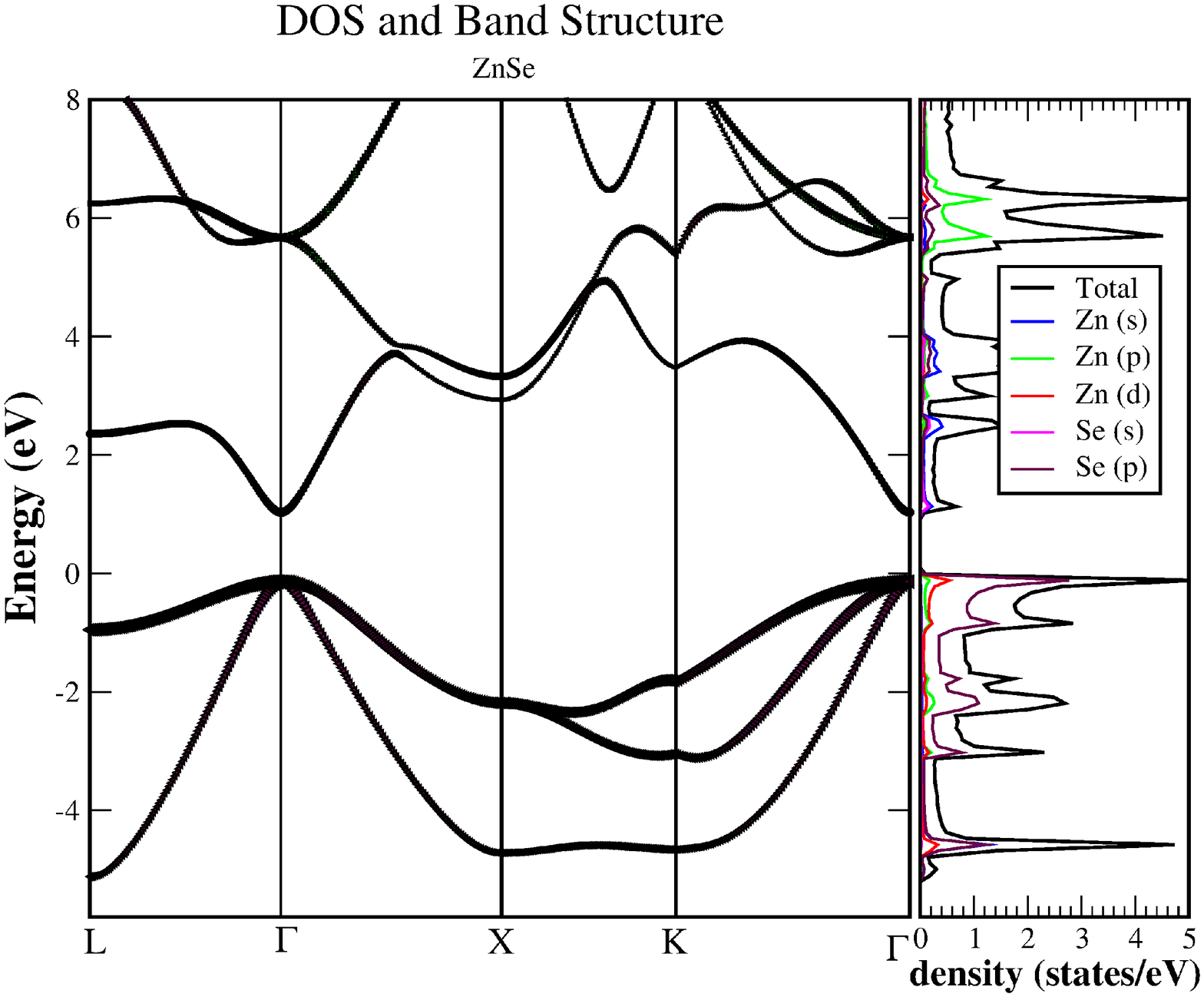} 
\caption{Band structure and density of states for ZnSe, Fermi level is set to zero at valence band maximum}
\label{fig1}
\end{figure}

The electronic structure calculations \cite{ref26} are performed using first-principles methods within the frame-work of DFT with Perdew-Burke Ernzerhof exchange correlation energy functional\cite{pbe} based on a generalized gradient approximation. We used a projector augmented wave method as implemented in Vienna \textit{ab-initio} simulation package (\textsl{VASP})\cite{ref8,ref9,ref10}. The Kohn-Sham wave functions of the valence electrons were expanded in plane wave basis with energy cut-off of 500 eV.  Ionic relaxation was performed using conjugate-gradient method, until forces were reduced to within 0.01 eV/Angstrom. The Brillouin zone sampling was carried out using Monkhorst Pack grid of 11x11x11 k-points. The band structure is computed along the high-symmetry k-points in the irreducible Brillouin zone, with 100 k-points between each pair of high-symmetry points. Computed band structure with the self-consistent density of states (DOS) is shown in the Fig.\ref{fig1}. Since as an input for the transport calculation within Rode's method, only band structure for one valley is needed, we have performed non-self consistent calculations of the band energies in a special k-point mesh around the $\Gamma$ point with 8531 k-points. Using such a dense mesh we have obtained very accurate group velocity and effective mass. 
\subsection{Scattering Mechanisms}
In this work, for calculating the mobility of ZnSe, four different types of scattering mechanisms are included (i) Ionized impurity scattering, (ii) Acoustic deformation potential scattering, (iii) Piezoelectric scattering and (iv) Polar optical phonon scattering. For ionized impurity scattering, the  formulation provided by the Brooks-Herring approach is used. The ionized impurity is a significant scattering mechanism at low temperatures and high doping concentrations. The momentum relaxation rate for this mechanism with its abinito counterpart is given by following expression \cite{ref1,ref7}
\begin{equation}
\frac{1}{\tau_{ii}(k)} = \frac{e^4 N }{8\pi\epsilon_0^2\hbar^2 k^2 v(k)}[D(k)ln(1+\frac{4k^2}{\beta^2})-B(k)] ,
\label{Ionized impurity}
\end{equation}
where $\epsilon_0$ is the dielectric constant, $\hbar$ is the reduced Planck constant, and $\beta$ is the inverse screening length given by
\begin{equation}
\beta^2 = \frac{e^2}{\epsilon_0 k_B T}\int D_s(\epsilon) f(1-f)d\epsilon,
\label{beta square}
\end{equation}
where N is the concentration of ionized impurity and it is given by
\begin{equation}
\ N = N_A + N_D
\label{impurity}
\end{equation} 
where $N_A$ and $N_D$ are the acceptor and donor concentrations respectively. The expressions for D(k) and B(k) are taken from equations (91) and (92) of Ref. \cite{ref7}. \\

\begin{equation}
\ D(k) = 1 + \frac{2 \beta^2 c^2}{k^2} + \frac{3 \beta^4 c^4}{4 k^4}
\label{D(k)}
\end{equation} 

\begin{equation}
\ B(k) = \frac{4 k^2/\beta^2}{1 + 4 k^2/\beta^2} + 8 \frac{\beta^2+2k^2}{\beta^2+4k^2}c^2 +   \frac{3\beta^4+6\beta^2 k^2 -8k^4}{(\beta^2+4k^2)k^2}c^4
\label{B(k)}
\end{equation} 

\indent Acoustic deformation occurs due to the coupling of electrons with non-polar acoustic phonons. The momentum relaxation rate for acoustic deformation potential scattering is given by \cite{ref1,ref7}
\begin{equation}
\frac{1}{\tau_{ac}(k)} = \frac{e^2 k_B T E_D^2 k^2}{3\pi\hbar^2 c_{el}v(k)}[3 - 8c^2(k)+6c^4(k)] ,
\label{acoustic deformation}
\end{equation}
where c(k) is the contribution of the p-type orbital to the wave function of the conduction band, $c_{el}$ is spherically averaged elastic constant and $E_D$ is acoustic deformation potential and is given by conduction band shift (in eV) per unit strain due to acoustic waves.  For \textit{ab-initio} calculations, the wave function admixture $c(k)$ is obtained through projecting the Kohn-Sham wavefunctions onto the spherical harmonics which are non-zero only within spheres centering the ions and is implemented in \textsl{VASP} package. \\
\indent Chemical bonds in compound semiconductors such as ZnSe are partly ionic in nature. Zn atom has a slight positive and Se atom has a slight negative charge. The magnitude of this charge is determined by the degree of the ionic nature of the bond, and it is a fraction of electronic charge \cite{ref24}. The vibrations of atoms cause changes in the lattice constant. This perturbs the dipole moment between the atoms that eventually scatter the electrons. The of polar scattering due to the long-wavelength acoustic phonons is called piezoelectric scattering and the polar scattering due to optical phonons is called polar optical phonon (POP) scattering. Piezoelectric scattering is important at low temperatures and at low doping densities in polar materials. Since ZnSe is polar, it necessary to include piezoelectric scattering to get a better fit for the mobility output at low temperature. The momentum relaxation rate for piezoelectric scattering with ab-inito parameters as input is given by \cite{ref1,ref7}
\begin{equation}
\frac{1}{\tau_{pz}(k)} = \frac{e^2 k_B T P^2 }{6\pi\epsilon_0\hbar^2 v(k)}[3 - 6c^2(k)+4c^4(k)] 
\label{pz}
\end{equation}
where $P$ is dimensionless piezoelectric coefficient and it is given by \cite{ref7}
\begin{equation}
\ P^2 = h_{14}^2\epsilon_0\frac{[(\frac{12}{c_l})+(\frac{16}{c_t})]}{35} 
\label{pzcoeff}
\end{equation}
where $h_{14}$ is one element of piezoelectric stress tensor and  $c_l$ $c_t$ are the spherically averaged elastic constant for longitudinal and transverse modes respectively and are given by \cite{ref7}
\begin{equation}
\ c_l = (3c_{11} + 2c_{12} + 4c_{44})/5 
\label{cl}
\end{equation}

\begin{equation}
\ c_t = (c_{11} - c_{12} + 3c_{44})/5 
\label{ct}
\end{equation}

where $c_{11}$, $c_{12}$ and $c_{44}$ are three independent elastic constants.
\quad
\newline The POP scattering is the most dominant scattering mechanism near room temperature and in the higher temperature regime. Since POP scattering is inelastic and anisotropic, the Rode iterative scheme is used to directly evaluate the scattering rates so that the momentum relaxation rate is given by \cite{ref5,ref7}
\begin{equation}
S_{o} = ( N_{po} + 1 - \textit{f}^- ) \lambda^{-}_0 + ( N_{po} + \textit{f}^+) \lambda^{+}_0 
\label{So}
\end{equation}
\begin{equation}
\lambda^+_{o} = \beta^+[(A^+)^2 ln \mid \frac{k^+ + k}{k^+ - k}\mid - A^+ c c^+ - aa^+cc^+] 
\label{lambda0}
\end{equation}
\begin{equation}
\beta^+ = \frac{e^2 \omega_{po} k^+ }{4 \pi \hbar k v(k^+)} (\frac{1}{\epsilon_\infty} - \frac{1}{\epsilon_0})
\label{betaaaa}
\end{equation}
\begin{equation}
A^+ = aa^+ + \frac{(k^+)^2 + k^2}{2 k^+ k} cc^+,
\label{Apositive}
\end{equation}
where the subscript plus denotes the scattering out by absorption so it is to be evaluated at an energy $\epsilon + \hbar \omega_{po} $ and the subscript minus denotes scattering out by emission so that it is to be evaluated at energy  $\epsilon - \hbar \omega_{po} $. If energy is less than $ \hbar \omega_{po} $, then the emission of phonons is not possible and hence $\lambda^{-}_0$ is to be considered to be zero, 
$N_{po} $ is the number of phonons and is given by 
\begin{equation}
N_{po} = \frac{1}{exp(\hbar \omega_{po}/k_B T)-1} .
\label{Npo}
\end{equation}
The in scattering operator for POP is given by \cite{ref5,ref7}
\begin{equation}
S_{i} = ( N_{po} + \textit{f} ) \lambda^{-}_i g^- + ( N_{po} + 1 - \textit{f}) \lambda^{+}_i g^+ 
\label{Si}
\end{equation}
\begin{equation}
\lambda^+_{i}(k) = \beta^+ [\frac{(k^+)^2+k^2}{2k^+k}(A^+)^2 ln\mid \frac{k^+ + k}{k^+ - k}\mid - (A^+)^2 - \frac{c^2 (c^+)^2}{3}]
\label{lambdai}
\end{equation}
\subsection{\textit{Ab-initio} Inputs}
\begin{figure}
\includegraphics[width=120mm,height=100mm]{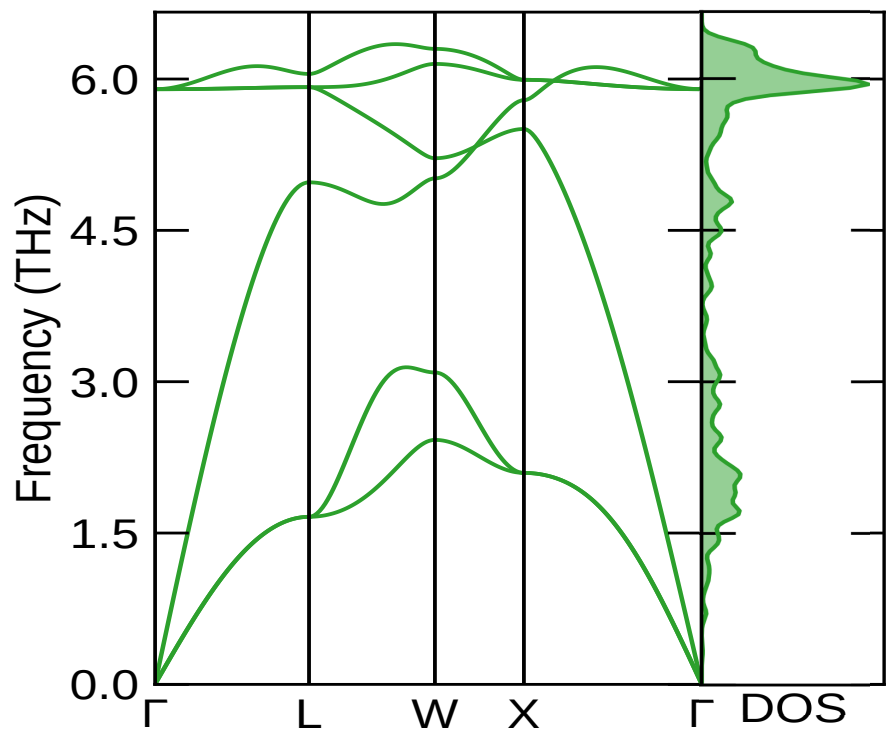}
\caption{Phonon band structure with phonon density of states for ZnSe }
\label{fig22}
\end{figure}

The computed band structure and density of states for ZnSe is shown in Fig.~\ref{fig1}.  As already mentioned, we have calculated the band structure using the density functional theory using a three dimensional $k$ mesh around the conduction band minimum (CBM). The conduction band is expressed as the average energy of the electrons as a function of k from the CBM.  In order to evaluate the group velocities, we have first calculated the  derivatives of the conduction band energy  with respect to $k$, performed an analytical fitting of the conduction band with a six degree polynomial and divided the conduction band into four segments \cite{ref1,ref11} to obtain a smooth curve for both mobility and conductivity. The Fermi level is obtained by calculating the carrier concentration using equation \eqref{fermi} and matching it to the given concentration: 
\begin{equation}
n = \frac{1}{V_0} \int_{\epsilon_c}^{\infty}{D_S(\epsilon)f(\epsilon) d\epsilon},
\label{fermi}
\end{equation}
where $D_S(\epsilon)$ represents density of states at energy $\epsilon$, where $\epsilon_c$ represents the bottom of conduction band and $V_0$ represents the volume of the cell.   
We have calculated the low and high frequency dielectric constants \cite{ref12, ref13}, polar optical phonon frequencies $\omega_{po}$ \cite{ref14}, the elastic constant \cite{ref24,ref25} and the piezoelectric constant by using the density functional perturbation theory as implemented in VASP. Calculated abintio value of low and high frequency dielectric constant is 7.45 and 3.44 respectively. We have obtained \textit{abinitio} polar optical phonon frequency of 5.88 THz. The computed phonon band structure with phonon density of states for ZnSe is shown in figure \ref{fig22}. The acoustic deformation potential is obtained by mimicking a uniform lattice deformation due to acoustic phonons which results a shift of the conduction band minimum. The deformation potential is given by
\begin{equation}
E_D=-V(\frac{\partial E_{CBM}}{\partial V})|_{V=V_0}
\label{ADP}
\end{equation}
Where E$_{CBM}$ is the conduction band minimum and V$_0$ is the equilibrium volume. We have obtained \textit{abinitio} value of 0.574 nm of lattice constant, 1.17 eV of band gap, 12 eV of acoustic deformation potential, 0.0392 of Piezoelectric coefficient and -0.08162 $C/m^2$ of Piezoelectric constant($e_{14}$). Calculated \textit{abinitio} value of elastic constant $c_{11}$, $c_{12}$ and $c_{44}$ are $7.99 \times 10^{10} N/m^2$, $4.54 \times 10^{10} N/m^2$ and $3.71 \times 10^{10} N/m^2$ respectively. 
\par
Electronic correlation effects often play a major role in the carrier transport for $\mathrm{3d-4p}$ systems. An efficient approach for the inclusion of such correlation effect  is multiband mean field Hubbard model (DFT+U). Karazhanov \textit{et al.}~\cite{ref27} used this approach to study the effect of correlation in ZnX (X = O, S, Se, Te) and concluded that the effects of correlation in ZnO are more important than in ZnS, ZnSe and ZnTe. The spin-orbit interaction (SOI) is another effect that we have not included in our calculation. Usually for the semiconductors with zinc blend structures such as ZnSe, the effect of SOI is to lift the six-fold degeneracy of the valence band edge at $\Gamma$-point. However, since we are only interested in electronic transport (not the hole) and our formalism does not depend on the exact position or the nature of the edge of the valence band, we have neglected the effect of the SOI.
  
\subsection{Simulation Flowchart}
\begin{figure}[H]
\includegraphics[width=180mm,height=200mm]{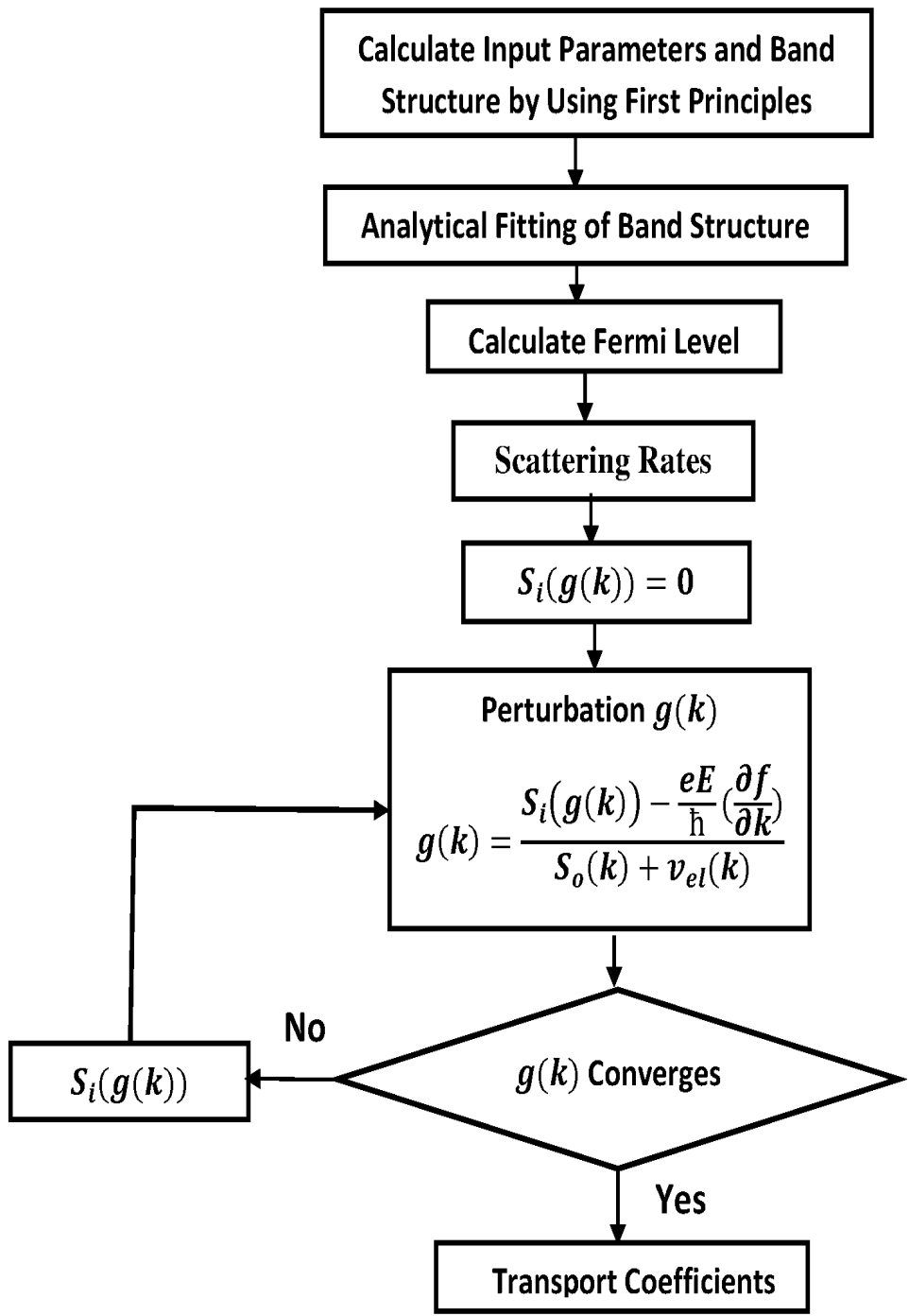}
\caption{ Flowchart for transport coefficient calculation from \textit{ab-initio} inputs}
\label{flowchart}
\end{figure}

The flowchart for our mobility and conductivity calculations using the Rode method with the \textit{ab-initio} inputs is shown in Fig.~\ref{flowchart}. First, we have to calculate all the required inputs using the first principles methods detailed in the previous section. Then we have to perform the analytical fitting of the band structure, to obtain smooth curves in order to calculate the group velocities of the carriers with $k$. Then the smoothed band structure is used to calculate the Fermi level using Eq. \eqref{fermi}. The calculations are sensitive with respect to the calculated Fermi level and hence must be done carefully with good precision. Next we have to calculate various scattering rates for different scattering mechanisms by using \eqref{Ionized impurity}, \eqref{acoustic deformation}, \eqref{pz}, and \eqref{So}. These scattering rates are then used to calculate the perturbation in the distribution function using \eqref{pert}. Now the obtained $g(k)$ is used to calculate the desired transport coefficients. If $g(k)$ is obtained right after the first order iterate, it will replicate the RTA results of the transport coefficients.   
\section{Results}
\subsection{Scattering Rate vs Energy}
\begin{figure}[H]
\subfigure[$ $ 30 K]{\includegraphics[width=75mm,height=50mm]{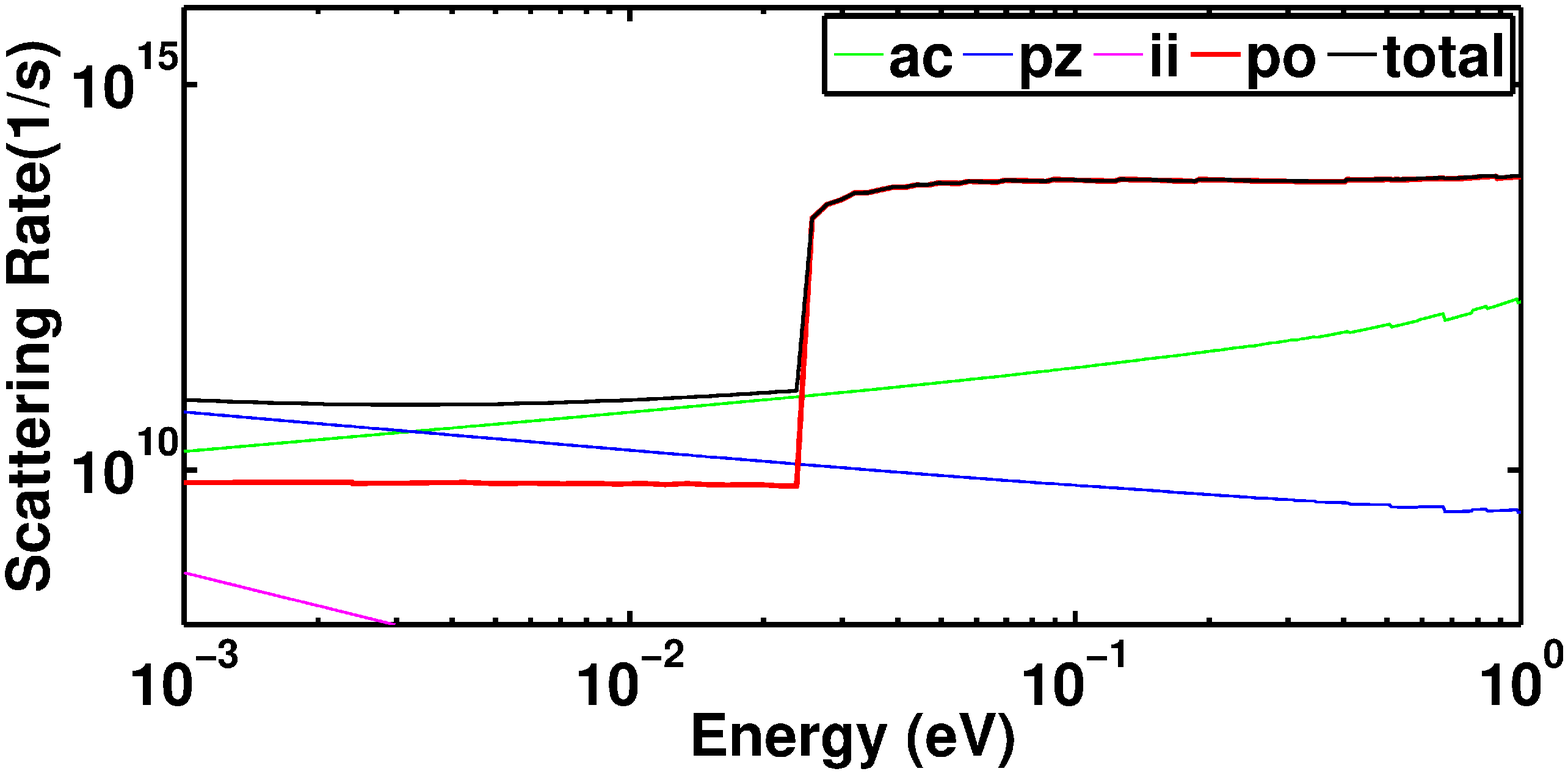} }
\subfigure[$ $ 300 K]{\includegraphics[width=75mm,height=50mm]{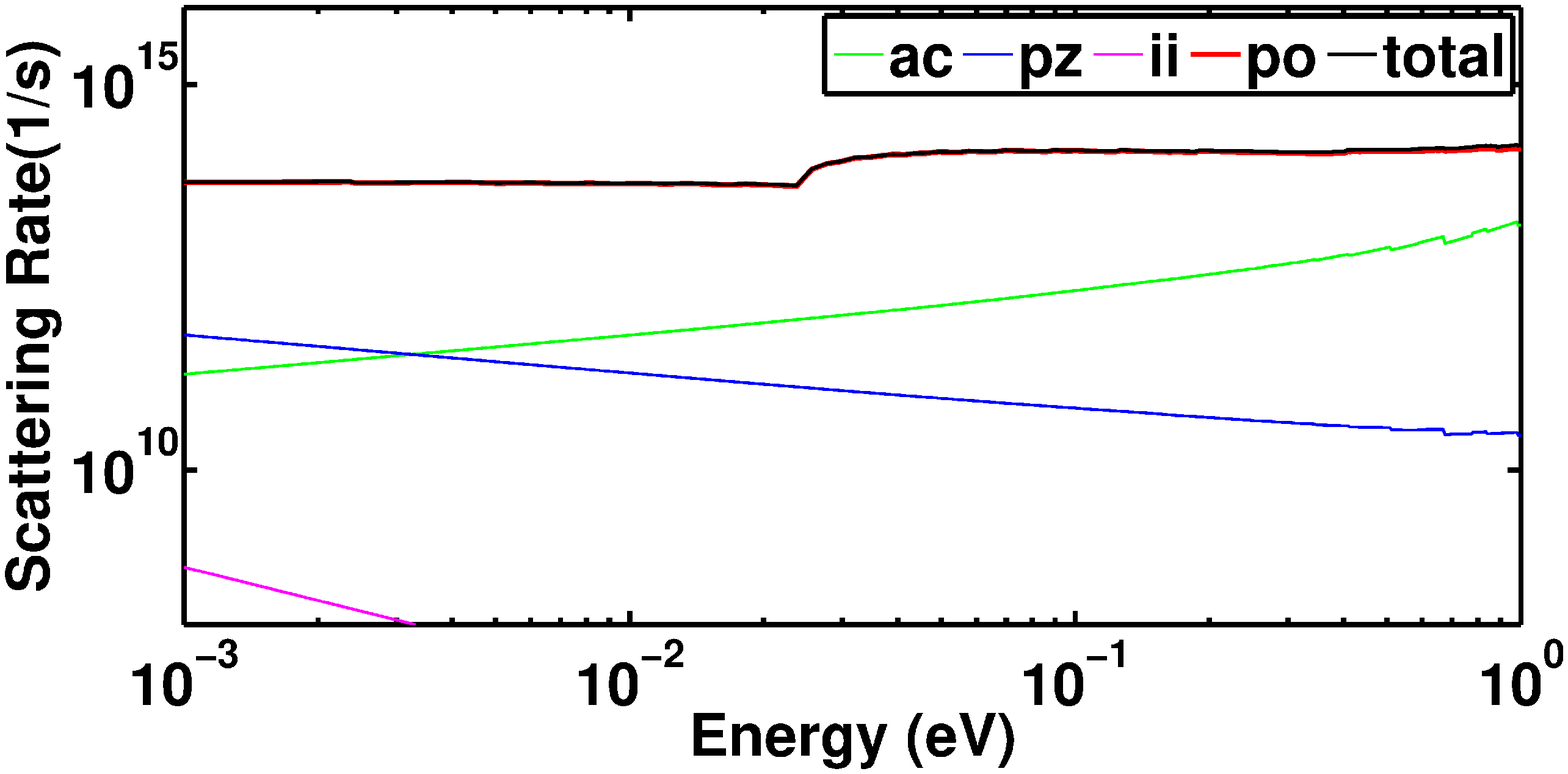} }

\caption{Scattering Rates vs Energy for $N_D = 1\times10^{10}cm^{-3}$ }
\label{fig2}
\end{figure}

\begin{figure}[H]
\subfigure[ $$ 30 K]{\includegraphics[width=75mm,height=50mm]{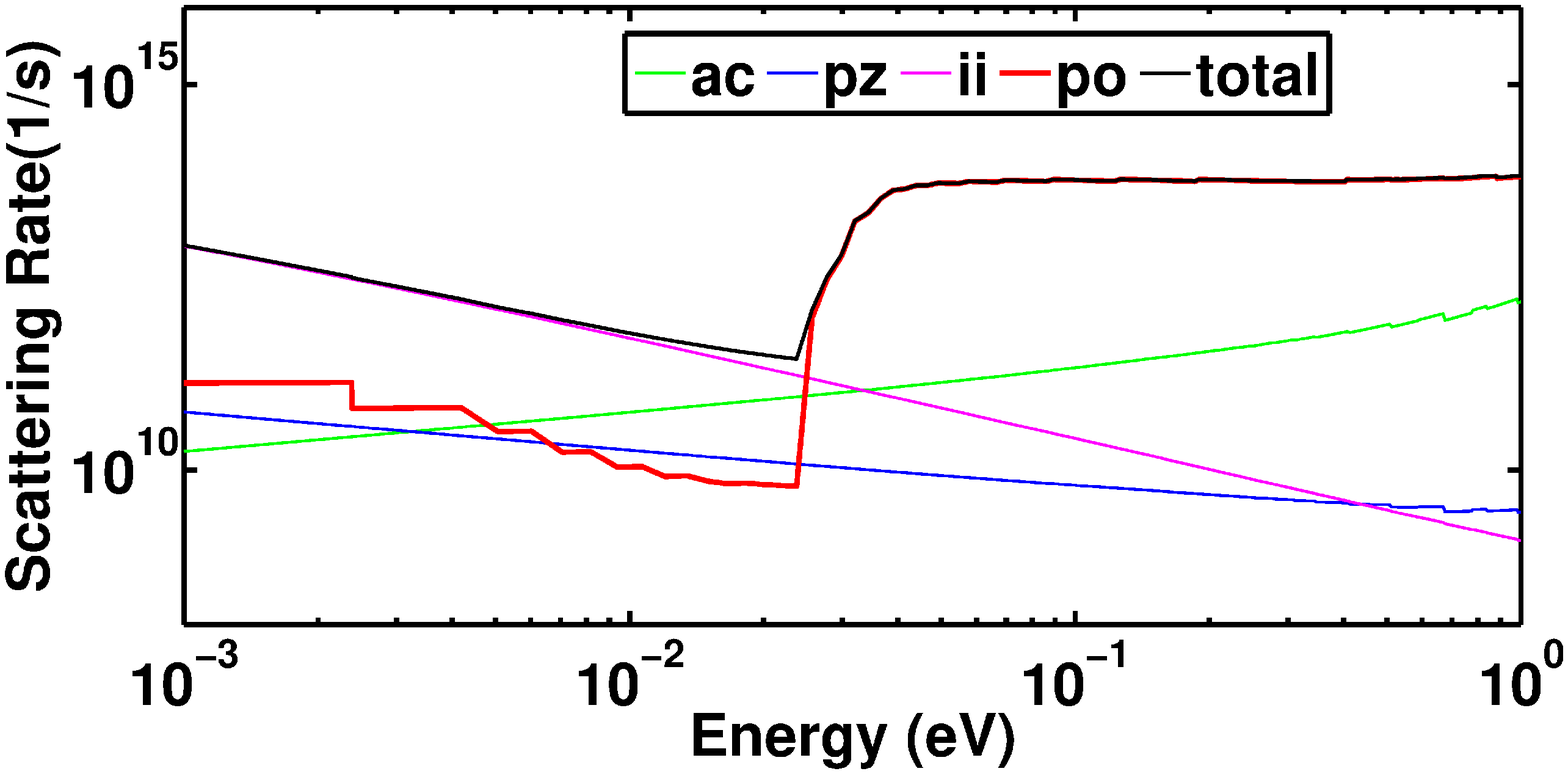} }
\subfigure[ $$ 300 K]{\includegraphics[width=75mm,height=50mm]{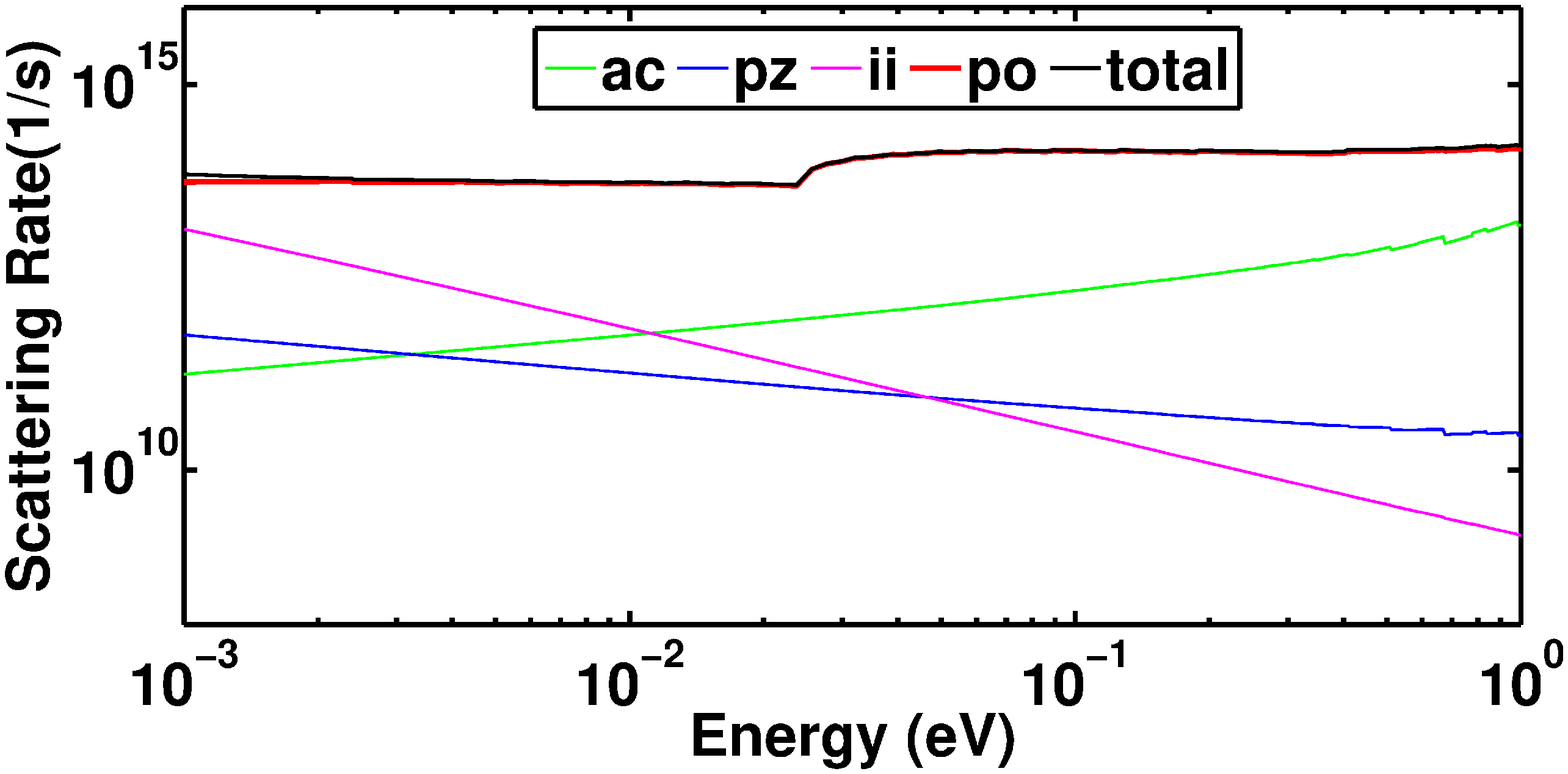} }
\caption{Scattering Rates vs Energy for $N_D=1\times10^{15}cm^{-3}$}
\label{fig3}
\end{figure}

\quad
\newline For a comparison between the different scattering mechanisms in n-ZnSe, the scattering rate as a function of energy is plotted in Fig.~\ref{fig2} and Fig.~\ref{fig3}, for different doping concentrations $ N_D = 1 \times 10^{10} cm^{-3} $ and $N_D = 1 \times 10^{15} cm^{-3} $ at temperatures 30 K and 300 K. At low doping and lower temperatures, piezoelectric scattering is considered the most dominant scattering mechanism for low energy carriers. At a temperature of 30 K, the average energy of carriers is $\frac{3}{2}k_B T = 0.0038 eV$, and hence most of the carriers are in the low energy region. At low temperatures, around the average energy of carriers, both piezoelectric and acoustic deformation potential scattering mechanisms are approximately equal. So at lower doping and at lower temperatures, piezoelectric scattering and acoustic deformation potential scattering are the most dominant scattering mechanisms for ZnSe. \\
 \indent At higher doping ($1\times 10^{15}$) for low temperatures, ionized impurity scattering is the most dominant for low energy carriers. In Fig.~\ref{fig2} and Fig.~\ref{fig3} there is a sudden change in POP scattering rate after a particular energy, which is due to the fact that if an electron energy is smaller than the POP energy $\hbar \omega_{po} = 0.024 eV$, then it can scatter only by the absorption of an optical phonon, but if the energy of an electron is larger than the optical phonon energy, then it can scatter by both emission and absorption of optical phonons. At 300 K, POP scattering is the most dominant scattering mechanism for all electron energies for both higher and lower doping making it the most important scattering mechanism at temperatures ranging from room temperature to higher temperature. Since all working devices operate in this temperature region, it is necessary to properly include POP scattering for a good theoretical mobility calculation. Since RTA is not valid for inelastic scattering mechanisms, other abinito RTA based codes such as the BoltzTrap \cite{ref20} code may not produce the desired results in the case of n-ZnSe.    
 \subsection{Electron Mobility}
\begin{figure}[H]
\includegraphics[width=150mm,height=120mm]{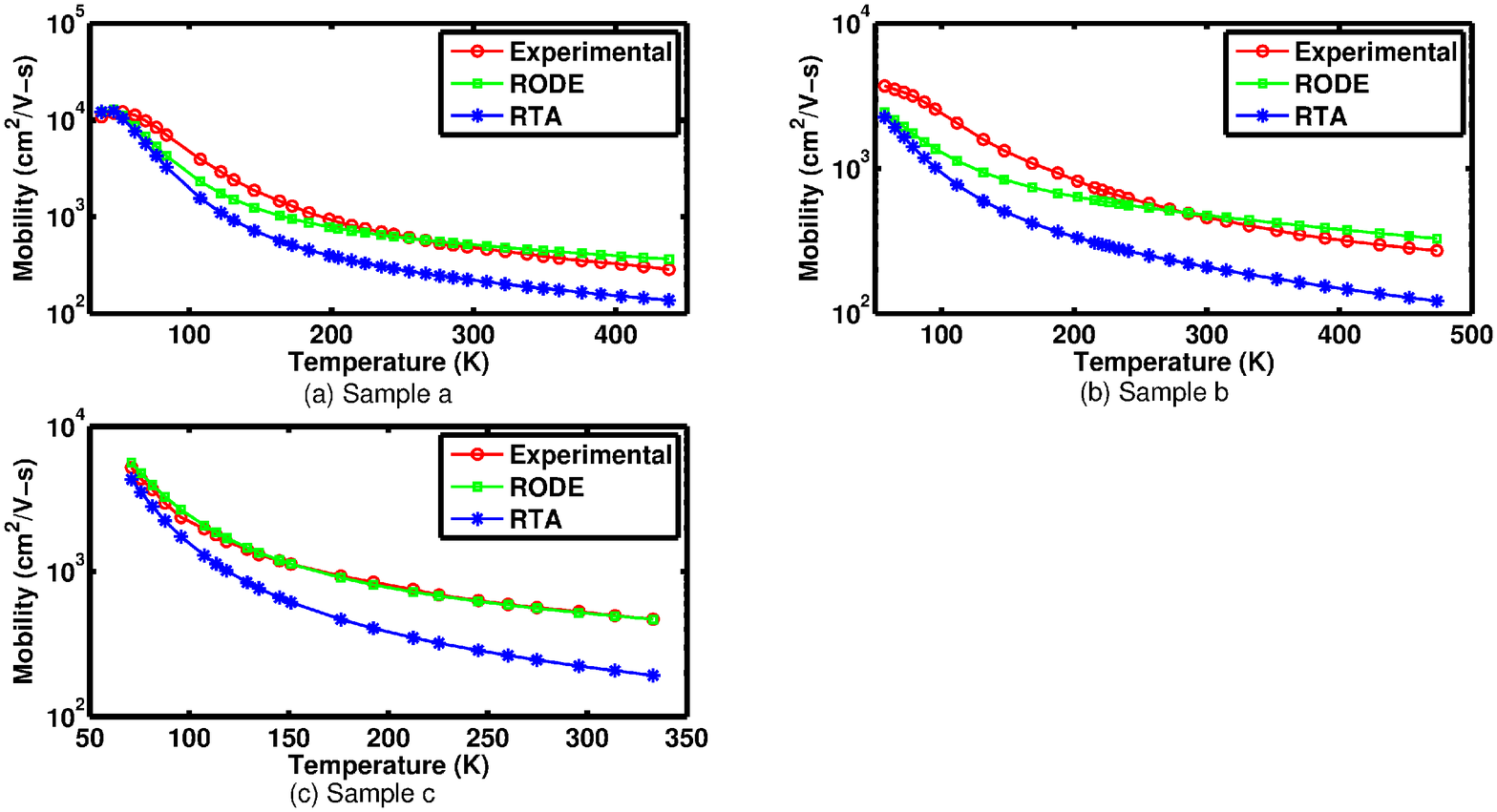} 
\caption{Calculated and experimental measured mobility with temperature variation for ZnSe at Different Doping. More detail about donor and acceptor concentration is shown in table \ref{table2}}
\label{fig4}
\end{figure}

\begin{figure} 
\includegraphics[width=100mm,height=80mm]{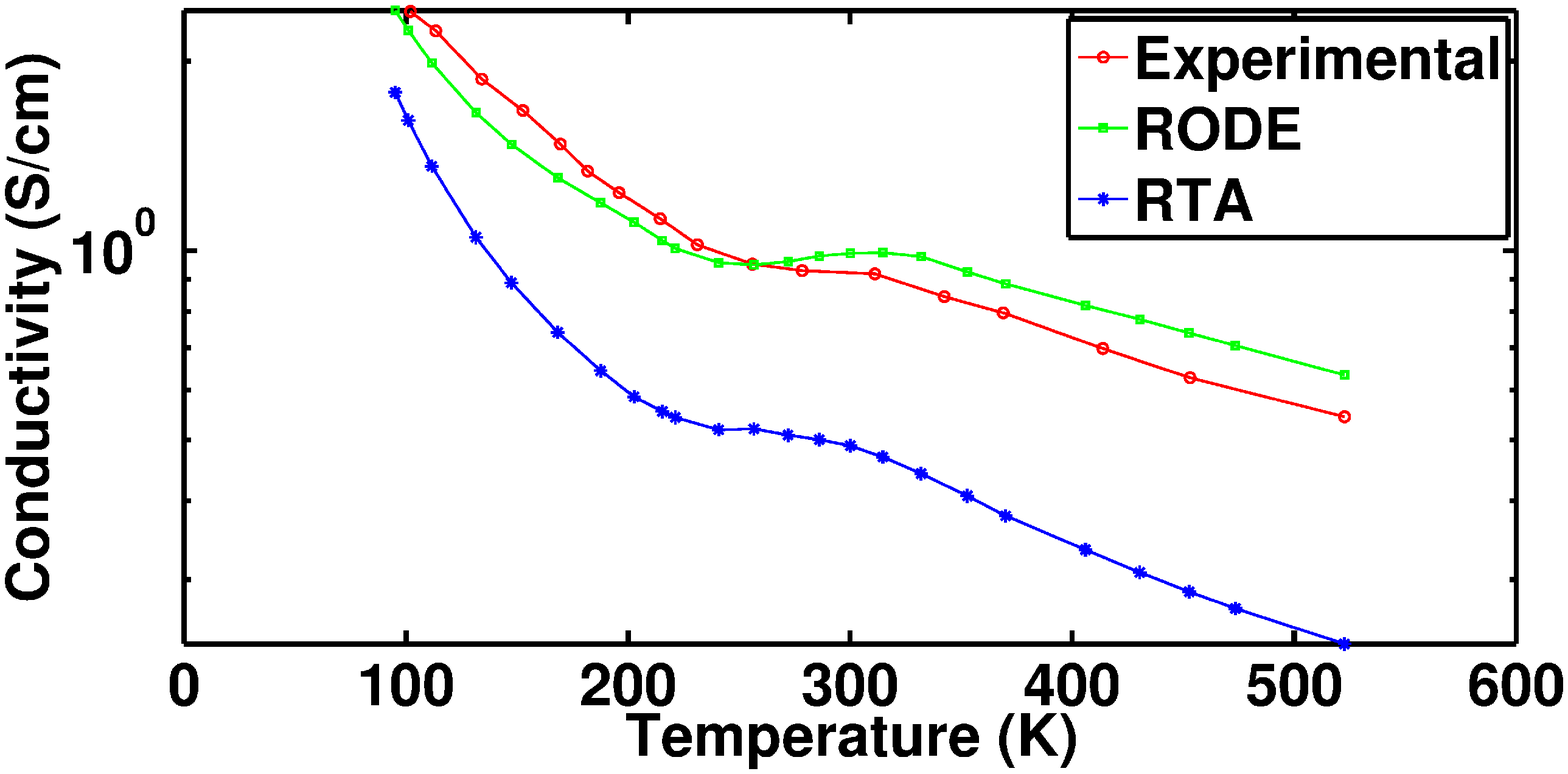}
\caption{Calculated and experimental measured conductivity with temperature variation for sample b}
\label{fig5}
\end{figure}

\begin{figure}
\includegraphics[width=100mm,height=80mm]{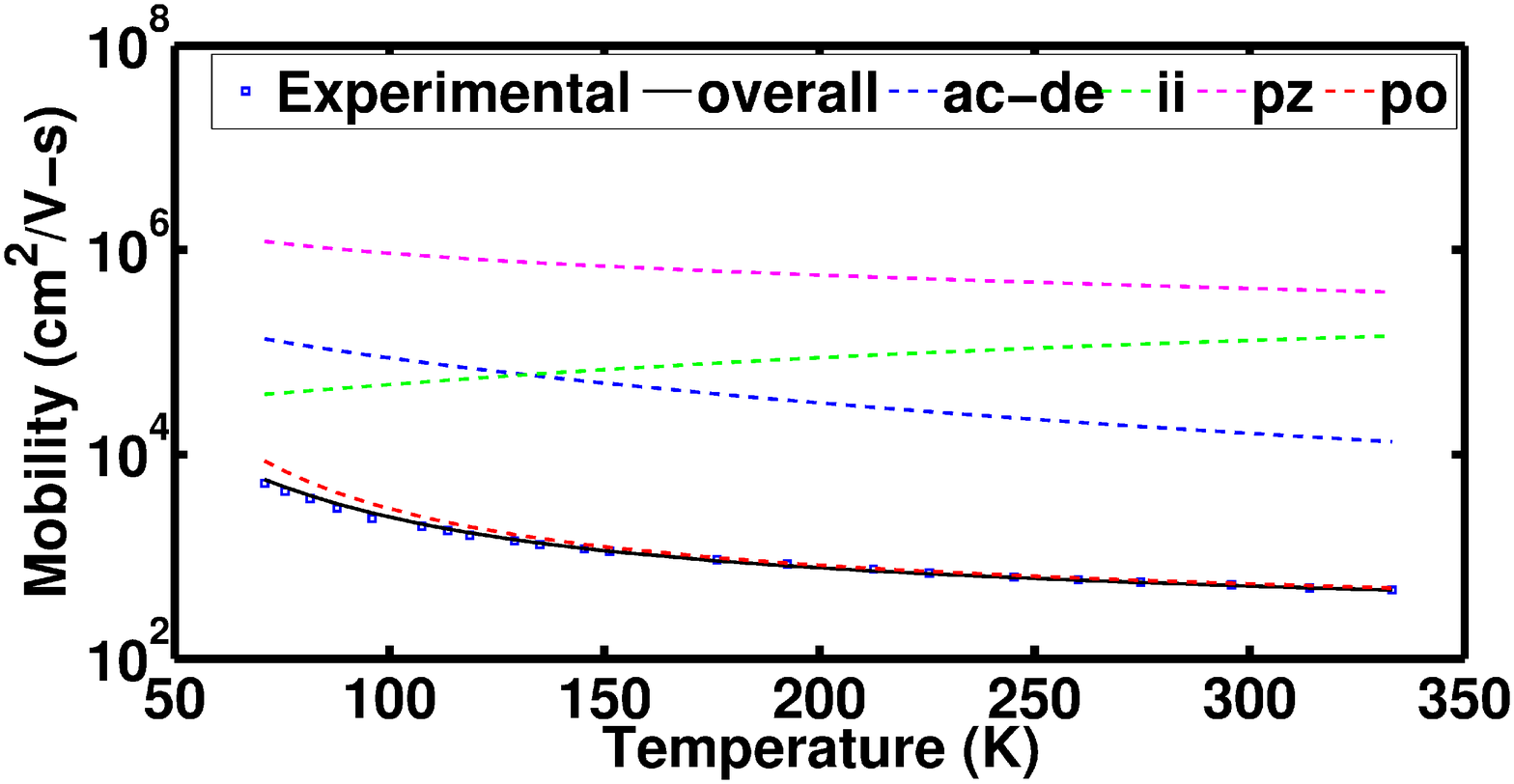}
\caption{Contribution of mobility from different scattering mechanisms for Sample (c)}
\label{fig6}
\end{figure}

\begin{figure}
\includegraphics[width=100mm,height=80mm]{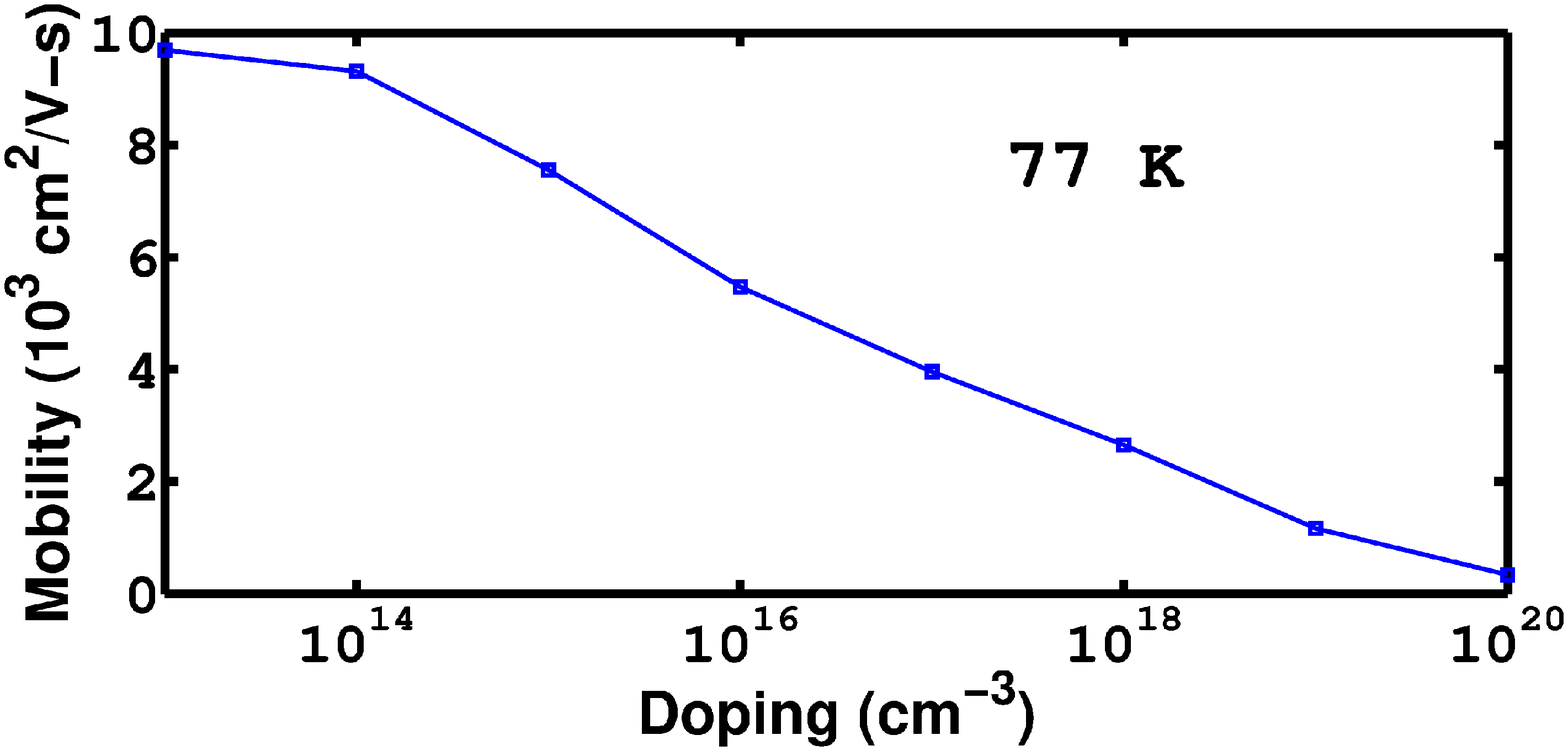}
\caption{Calculated mobility for different doping concentration at 77 K}
\label{fig7}
\end{figure}

\begin{figure}
\includegraphics[width=100mm,height=80mm]{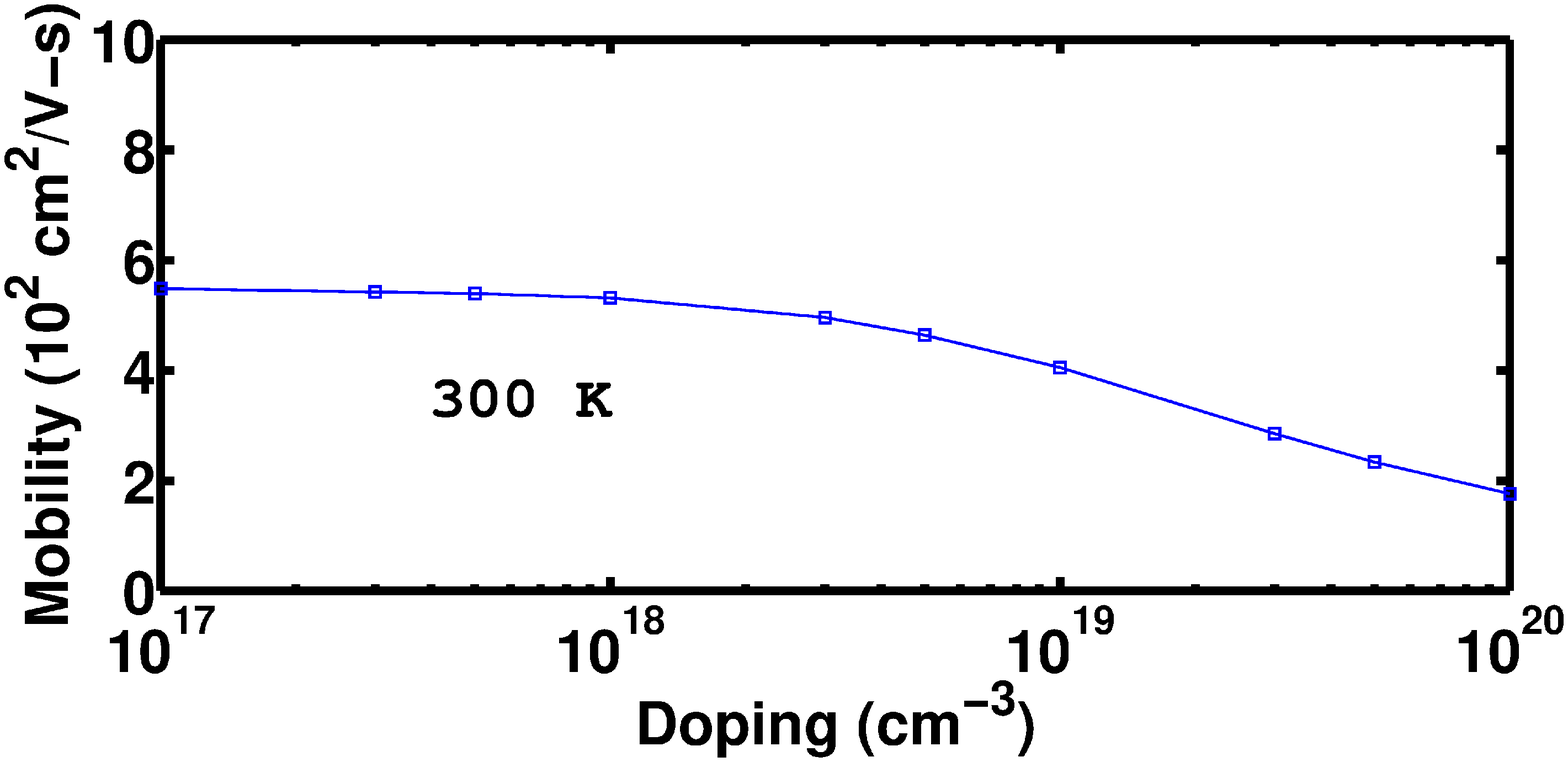}
\caption{Calculated mobility for different doping concentration at 300 K}
\label{fig8}
\end{figure}

By using the Rode iterative method based on the flowchart detailed above, the electron mobility is calculated for three different experimentally characterized n-ZnSe samples of different doping concentrations over a wide temperature range. The donor and acceptor concentrations for different experimentally characterized n-ZnSe samples taken from different references is shown in Tab.~\ref{table2}. Figure \ref{fig4} shows  a comparison between the experimentally measured and theoretically calculated mobility using our Rode scheme as well as the RTA method. Overall there is good agreement between the theoretical and experimental curves when considering the Rode scheme results and score much better than RTA results. The RTA underestimates the mobility for all samples, which is due to POP scattering, since POP scattering is inelastic and anisotropic and hence changes the distribution function such that a constant or simple power law form for the relaxation time cannot be defined. The calculated mobility for sample (a), (b) and (c) using the Rode method have an average relative error of $ 18.44\%$ and $21.63\% $ and $4.08\%$ respectively. While the mobility calculated using the RTA have average errors of $ 113.46\%$, $ 55.69\%$ and $ 42.77\%$ for sample (a), (b) and (c) respectively.\\
\indent For sample (a), the Rode calculation shows an error of $ -15.03\%$ and $ -7.27\%$ at 200 K and 300 K respectively, while the RTA shows an error of $ -133.45\%$ and $ -117.88\% $ at 200 K and 300 K respectively. At lower temperatures, even RTA shows a smaller error since at lower temperatures, POP scattering becomes insignificant, since the POP energy of $\hbar \omega_{po} = 0.024 eV$ is required. Around 40 K or lower, the electrons have smaller probability of scattering by the absorption or emission of 0.024 eV of optical phonon energy. The Rode method shows an error of $ -6.45\%$ at 300 K for sample (b), while the RTA shows an error of $-54.26\%$ at 300 K with respect to the experimental curve. For sample (c), the Rode calculation shows an error of $ -3.86\%$ and $-1.05\%$ at 200 K and 300 K respectively, while the RTA shows an error of $ -51.49\% $ and $ -57.85\% $ at 200 K and 300 K respectively. Therefore, all in all, the RTA results are quite inappropriate for mobility calculations of polar materials. \\
\indent Figure \ref{fig5} shows the calculated conductivity with respect to the experimental one for sample (b). Sample (b) contains shallow as well as deep donors. With the increasing temperature conductivity deceases in experimental curve, but around 230 K there is sudden change in conductivity curve slope, this is due to ionization of deep donors. Deep donors have a doping concentration of $ N_{deep} = 4.5 \times 10^{15} cm^{-3}$ \cite{ref22} and have ionization energy of $ E_{deep} = 130 meV $ \cite{ref22}. The temperature dependent free electron concentration for the semiconductor having deep donors is given by equation

\begin{equation}
\frac{n(n-n_0)}{N_{deep} + n_0 - n}  = \frac{N_C}{g} exp(\frac{-E_{deep}}{k_BT}) 
\label{deep}
\end{equation}
where $g$ is the degeneracy of deeper level, here $g = 2$ is taken for calculation and $n_0$ is the concentration of the electrons activated from shallow donors in the region of their exhaustion. 
The conductivity calculated by the Rode iterative method is again much better than that calculated by using the RTA method. The calculated conductivity shows a very good qualitative and quantitative agreement with respect to the experimental data. Figure \ref{fig6} shows the overall mobility and the mobility by considering only one type of scattering mechanism at a time for sample (c). The lower mobility is most dominant in deciding the overall mobility. From this figure it is clear that POP is the most dominant scattering mechanism, from 70 K to higher temperatures. Figures \ref{fig7} and \ref{fig8} show the mobility for different doping concentrations at 77 K and 300 K by assuming a compensation ratio of unity. With increasing doping concentration, the curves show a decrease in mobility due to an increase in the number of ionized centers. 
\section{Conclusion}
We presented an \textit{ab-inito} semi-classical transport calculation for the mobility and conductivity of n-ZnSe by using Rode iterative method in order to conclusively illustrate the role of inelastic scattering processes.  We recognized that inelastic polar optical phonon scattering is the most dominant scattering mechanism over most of the temperature region.   A good agreement with various experiments was observed for different doping concentrations over a wider range of temperatures. In comparing our results against that obtained using the relaxation time approximation method, we clearly noted the discrepancy in explaining experimental results, thereby pointing out the need to advance semi-classical transport calculations beyond the relaxation time approximation. Further work will be extended p-type semiconductor and multivalley transport.


\newpage 

\clearpage
\newpage
\clearpage
\newpage
\newpage
\clearpage
\newpage
\clearpage
\newpage
\begin{table} [H]
\caption{ Doping of different experimentally fabricated n-ZnSe Samples}
\label{table2}
\begin{ruledtabular}
\begin{tabular}{cccccccc}
Sample  &  $N_D - N_A (cm^{-3})$ & Donor, $N_D(cm^{-3})$  & Acceptor, $ N_A(cm^{-3})$    \\
\hline                                                    
a \cite{ref21}      &   $1 \times 10^{15}$   & $2.9 \times 10^{15}$   & $1.9 \times 10^{15}$ \\
b \cite{ref22}      &   $1.1 \times 10^{16}$   & $6 \times 10^{16}$     & $4.9 \times 10^{16}$    \\
c \cite{ref23}      &   $6.3 \times 10^{15}$   & $7.5 \times 10^{15}$     & $1.2 \times 10^{15}$     \\
\end{tabular}
\end{ruledtabular}
\end{table}
\quad
\clearpage
\newpage
\clearpage
\newpage
\clearpage
\newpage
\clearpage
\newpage
\clearpage
\newpage


\begin{thebibliography}{99}
\bibitem{ref5} D. L. Rode, Phys. Rev. B \textbf{2}, 1012(1970).

\bibitem{ref6} D. L. Rode, Phys. Rev. B \textbf{2}, 4036(1970).

\bibitem{ref7} D. L. Rode, \textit{Semiconductors and Semimetals} (Academic Press, New York, 1975), Chapter 1.

\bibitem{refa} A. T. Ramu, L. E. Cassels, N. H. Hackman, H. Lu, J. M. O. Zide, and J. E. Bowers, Journal of Applied physics \textbf{107}, 083707 (2010).

\bibitem{refb} A. T. Ramu, L. E. Cassels, N. H. Hackman, H. Lu, J. M. O. Zide, and J. E. Bowers \textbf{109}, 033704 (2011).

\bibitem{refc} D. J. Howarth and E. H. Sondheimer, Proc. R. Soc. Lond. A \textbf{219}, 53(1953).
\bibitem{refd} H. Ehrenreich, Phys. Rev. \textbf{120}, 1951 (1960).
\bibitem{refe} H. E. Ruda and B. Lai, Journal of Applied physics \textbf{68}, 1714 (1990).
\bibitem{reff} H. E. Ruda, Journal of Applied Physics \textbf{59}, 1220(1986).

\bibitem{refg} A. Sztein, J. Haberstroh, J. E. Bowers, S. P. DenBaars, and S. Nakamura, Journal of Applied Physics \textbf{113}, 183707 (2013).

\bibitem{refh} A. Popescu, A. Datta, G. S. Nolas, and L. M. Woods, Journal OF Applied Physics \textbf{109}, 103709 (2011).

\bibitem{ref4} M. Lundstrom, \textit{Fundamentals of Carrier Transport}, 2nd ed.
(Cambridge University Press, Cambridge, UK, 2009).

\bibitem{refi} M. Aven and B. Segall, Phys. Rev. \textbf{130}, 81 (1963). 

\bibitem{refj} D. D. Nedeog, Phys. stat. sol.(b) \textbf{80}, 369(1977).

\bibitem{refk} 0. V. Emelyanenko, G.N.Ivanova, T.S. Lagunova, D. D. Nedeoglo, G. M. Shmelev and A. V. Simashkevich, phys. stat. sol. (b) \textbf{96}, 823(1979). 

\bibitem{refl} Y. Fukuda and M. Fukai, J.Phys. Soc. Japan \textbf{23}, 902(1967).

\bibitem{refm} B. R. Sethi, Phys. stat. sol. (a) \textbf{42}, 791 (1977).

\bibitem{ref1} A. Faghaninia, J. W. Ager III and C. S. Lo, Phys. Rev. B \textbf{91}, 235123 (2015).

\bibitem{ref2} Jasprit Singh, \textit{Electronic and Optoelectronic Properties of Semiconductor Structure}, (Cambridge University Press, Cambridge, UK, 2003).  

\bibitem{ref3} D. K. Ferry, \textit{Semiconductor Transport}, (Taylor \& Francis, London, 2000).

\bibitem{ref26} Y.Petroff, M.Balkanski, John P. Walter and Marvin L.Cohen, Solid State Communications \textbf{7}, 459(1969).

\bibitem{pbe}Perdew,~J.P, Burke,~K., Ernzerhof,~M., Physical Review Letters {\bf 77}, 3865 (1996)
\bibitem{ref8} G. Kresse and J. Hafner, Physical Review B \textbf{49}, 14251 (1994)
\bibitem{ref9} G. Kresse and J. Furthmüller, Physical Review B \textbf{54}, 11169 (1996).
\bibitem{ref10} G.Kresse and J. Furthmullerb, Computational Materials Science \textbf{6}, 15 (1996).

\bibitem{ref24} B. K. Ridley, \textit{Quantum processes in semiconductors}, (Clarendon Press, Oxford,1999).

\bibitem{ref11} A. Faghaninia, J. W. Ager and C. S. Lo, 2015 IEEE 42nd Photovoltaic Specialist Conference (PVSC), New Orleans, LA, 2015, pp. 1-4.
%

\bibitem{ref12} X. Gonze and Changyol Lee, Phys. Rev. B \textbf{55} 10355 (1997). 
\bibitem{ref13} B. Liu, M. Gu, Z. Qi, X. Liu, S. Huang and C. Ni, Phys. Rev. B \textbf{76}, 064307 (2007).  

\bibitem{ref14} P. Giannozzi and S. D. Gironcoli, Phys. Rev. B \textbf{43}, 7231 (1990).

\bibitem{ref24} B. H. Lee, Journal of Appl. Phys. \textbf{41}, 2984(1970).

\bibitem{ref25} C. G. Hodgins and J. C. Irwin, Physica Status Solidi (a) \textbf{28}, 647(1975).

\bibitem{ref27} A. Kartsev, D. Feya Oleg, N. Bondarenko and A. G. Kvashnin, Phys. Chem. Chem. Phys. \textbf{21}, 5262(2019).


\bibitem{ravindran} Karazhanov \textit{et al.}, JOURNAL OF APPLIED PHYSICS {\bf 100}, 043709 (2006)

\bibitem{ref20}  G. K. H Madsena and D. J. Singh, Computer Phy. Comm. \textbf{175}, 67(2006).

\bibitem{ref15} R.J. Nelmes, M.I. McMahon, \textit{Semiconductors and Semimetals} (Academic Press, New York, 1998), Chapter 3.

\bibitem{ref16} K. C. Agarwal, B. Daniel, T. Hofmann, M. Schubert, C. Klingshirn, M. Hetterich, Phys. Status Solidi B. \textbf{243}, 914 (2006). 

\bibitem{ref17} D. T. F. Marple, Journal Of Applied Physics \textbf{35}, 1879 (1964).

\bibitem{ref18} C. G. Hodgins and J. C. Irwin, Phys. Stat. Sol. \textbf{28}, 647 (1975). 	 
\bibitem{ref19} J. D. Zook, Phys. Rev. \textbf{136}, A869 (1964).

\bibitem{ref21} M. Aven, Journal of Applied Physics \textbf{42}, 1204(1971).

\bibitem{ref22} N. D. Nedeoglo, V. P. Sirkeli, D. D. Nedeoglo, R Laiho and E Lahderanta, Journal of Phy.: Condensed Matter \textbf{18}, 8113(2006).

\bibitem{ref23} A. N. Avdonin, D. D. Nedeoglo, N. D. Nedeoglo, and V. P. Sirkeli, Phys. Stat. Sol. (b) \textbf{238}, 45(2003).

\end{thebibliography}
\end{document}